\documentclass[12pt]{article}
\usepackage{amsmath}
\usepackage{amssymb}
\usepackage[dvips]{graphicx}
\usepackage[cp866]{inputenc}
\usepackage[english]{babel}

\newcommand{\beq}{\begin{eqnarray}}
\newcommand{\eeq}{\end{eqnarray}}
\newcommand{\be}{\begin{equation}}
\newcommand{\ee}{\end{equation}}

\def\fun#1#2{\lower3.6pt\vbox{\baselineskip0pt\lineskip.9pt
\ialign{$\mathsurround=0pt#1\hfil ##\hfil$\crcr#2\crcr\sim\crcr}}}

\newcommand{{\SD}}{\rm SD}

\newcommand{\vex}{\mbox{\boldmath${\rm x}$}}

\newcommand{\ver}{\mbox{\boldmath${\rm r}$}}

\newcommand{\vep}{\mbox{\boldmath${\rm p}$}}

\newcommand{\vez}{\mbox{\boldmath${\rm z}$}}

\newcommand{{\tr}}{{\rm tr}}

\newcommand{\lan}{\langle}
\newcommand{\ran}{\rangle}

\title{
\vspace{-15mm}
\rightline{\small ITEP--PH--1--2006}
\vspace{10mm}
\bf New nonperturbative approach to the  Debye mass in hot QCD}
\author{
N.~O.~Agasian\thanks{e-mail: agasian@itep.ru}~~and
Yu.~A.~Simonov\thanks{e-mail: simonov@itep.ru}
\\
{\textrm{State Research Center}}\\
{\textrm{Institute of Theoretical and Experimental Physics}} \\
{\textrm 117218 Moscow, Russia}}
\date{}

\begin{document}

\maketitle

\begin{abstract}
The  Debye mass $m_D$ is computed nonperturbatively in the
deconfined phase of QCD, where chromomagnetic confinement is known
to be present. The latter defines $m_D$   to be
$m_D=c_D\sqrt{\sigma_s}$, where $c_D \cong 2.06$ and
$\sigma_s=\sigma_s(T)$ is the spatial string tension.
The resulting magnitude of $m_D(T)$ and temperature dependence are
in good agreement with  lattice calculations.  Background
perturbation theory expansion for $m_D(T)$ is discussed in
comparison to standard perturbative results and recent
gauge-invariant definitions.
\end{abstract}

\vspace{1cm}


\section{Introduction}

The screening of electric fields in QCD was originally considered in
analogy to QED plasma, where  the Debye screening mass was well
understood \cite{1sc}, and the perturbative leading order (LO)
result for QCD was obtained long ago \cite{2sc}, $m_D^{\rm(LO)}
=\left( \frac{N_c}{3} + \frac{N_f}{6}\right)^{1/2} gT$. For not very
large $T$, however, the purely perturbative expansion is not
reliable, and attempts have been made to use the effective $3d$
theory \cite{3sc,4sc,5sc,6sc} to define the Debye mass $m_D$ through the
coefficients, which are to be  determined nonperturbatively
\cite{7sc}. In doing so one obtains a series \cite{7sc}, with the
leading term of the same form as $m_D^{\rm(LO)}.$

The lattice calculations of $m_D(T)$ have been made repeatedly
\cite{8sc}-\cite{15sc}, and  recently $m_D(T)$ was computed on the
lattice for $N_f=0,2$ \cite{13sc,14sc} using the free-energy
asymptotics
\be
\delta F_1(r,T)\equiv F_1(r,T) -F_1(\infty, T)
\approx -\frac43 \frac{\alpha_s(T)}{r} e^{-m_D(T)r},
\label{1sc}
\ee
where $F_1(r,T)$ was found from color singlet Polyakov
loop correlator. A comparison of lattice defined  $m_D(T)$ with $m_D^{\rm(LO)}$
made in \cite{15sc} in the interval from $T_c$ up to temperatures about $5.5 T_c$
shows that one requires a multiplicative
coefficient $A_{N_f=0} = 1.51$, $A_{N_f=2} = 1.42$.
A difficulty of the perturbative approach is that
the gauge-invariant definition of the one-gluon Debye mass is not available.
The purpose of our paper is to provide a gauge-invariant and
a nonperturbative method, which allows to obtain Debye masses in a
rather simple analytic calculational scheme.
In what follows we use the basically nonperturbative
approach of Field Correlator Method (FCM) \cite{16sc}-\cite{22sc}
and Background Perturbation Theory (BPTh) for nonzero $T$
\cite{23sc,24sc,25sc} to calculate $m_D(T)$ in a series, where the
first and dominant term is purely nonperturbative,
\be
m_D(T)=M_0 + {\rm BPTh~ series}.
\label{2sc}
\ee

Here $M_0$ is the gluelump mass due to chromomagnetic
confinement in  3d,  which is computed  to be $M_0 = c_D \sqrt{\sigma_s}$,
with $\sigma_s(T)$ being the spatial string tension and $c_D \simeq 2.06$ for $N_c=3$.
The latter is simply expressed in FCM through chromomagnetic correlator \cite{19sc},
and can be found either from  lattice measurements of the correlator itself  as in
\cite{21sc}, or from the 3d effective theory
\cite{3sc}-\cite{7sc}, $\sqrt{\sigma_s}=c_{\sigma} g^2(T)T$, or else from
the  lattice data \cite{26sc}. Therefore  $M_0(T)$ is
predicted for all $T$ and can be compared with lattice data
\cite{13sc,14sc,15sc}, see Fig.~\ref{fig_mdeb}.

We note, that $m_D(T)$ is defined here as the screening mass in
the static $Q\bar Q$ potential  $V_1$, which can be expressed
through the gauge-invariant correlator of  chromomagnetic and chromoelectric  fields
 \cite{21sc, 27sc,28sc}.  The screened Coulomb part of  the
 potential  $V_1$ coincides
with the singlet free energy $F_1 (r,T)$  at the   leading order
\cite{29a}, and in what follows we shall consider also the leading
order in BPTh, where the static potential $V_1(r)$ has a term of
the same form as the r.h.s. of Eq. (\ref{1sc}).

The paper is organized as follows. In section 2 the nonperturbative part and the
perturbative BPTh series for the thermal Wilson loop are defined, and the gluelump Greens
function is identified, using the path-integral formalism.
In section 3 an effective Hamiltonian is derived and the first
 terms of expansion (\ref{2sc}) are obtained for
  $m_D(T)$  computed through the spatial
string tension $\sigma_s(T)$. In section 4  a comparison is made  of $m_D(T)$
with lattice data and other approaches.
Section 5 is devoted to a short summary of results and outlook.

\section{Background Perturbation Theory for the thermal Wilson loop}

It is well
known that the introduction of the temperature for the quantum field system in
thermodynamic equilibrium is equivalent to compactification along the euclidean
"time" component $x_4$ with the radius $\beta=1/T$ and imposing
the periodic boundary conditions (PBC)
for boson fields (anti-periodic for fermion ones).
Thermal vacuum averages are defined in a standard way

\begin{equation}
\langle \ldots \rangle = \frac{1}{Z_{\beta}}
\int_{\rm PBC} [D A] \ldots e^{-S_{\beta}[A]},
\label{eq_vac}
\end{equation}
where partition function is

\be
Z_{\beta}=\int_{\rm PBC} [D A] e^{-S_{\beta}[A]},
\quad S_{\beta}=\int_0^{\beta} dx_4 \int d^3 x L_{YM}.
\label{eqZ}
\ee
One starts as in \cite{28sc} and \cite{30sc} with the correlator of Polyakov loops $\lan L(0) L^+ (\ver)\ran$

$$L(\vex) \equiv \frac{1}{N_c} \tr P\exp (ig \int^\beta_0 A_4(\vex, x_4) d x_4),$$
and obtains (cf. \cite{30sc}).

\be
\lan L(0) L^+ (\ver)\ran = \frac{1}{N^2_c} \exp \left( -\beta F_1(r, \beta)\right)
+ \frac{N^2_c-1}{N^2_c} \exp \left( -\beta F_8(r, \beta)\right).
\label{4sc}
\ee

As it is explained in the Appendix  in \cite{27sc}
 the representation (\ref{4sc}) can be obtained from two Polyakov
loops by identical deformation of contours with  tentackles
meeting at some  intermediate point and subsequent merging of
contour into one Wilson loop using completeness relation at the
meeting point  $\delta_{\alpha_1\beta_1} \delta_{\alpha_2 \beta_2}
= \frac{1}{N_c} \delta_{\alpha_1\beta_2} \delta_{\alpha_2 \beta_1}
+ 2 t^a_{\beta_2\alpha_1} t^a_{\beta_1\alpha_2}$, the first term
contributing to the free energy $F_1$ of the static $Q\bar Q$-pair in the singlet color state,
the second to the octet free energy $F_8$.
Accordingly one ends
for $F_1$ with the thermal Wilson loop of time extension $\beta =1/T$ and
space extension $r$,
\be
\exp \left( -\beta F_1(r, \beta)\right) = \lan W (r, \beta)\ran = \frac{1}{N_c}
\lan \tr P \exp (ig \int_C A_\mu dz_\mu) \ran.
\label{5sc}
\ee
Note, that in contrast to the case of the zero-temperature Wilson loop, the
averaging in (\ref{5sc}) is done with PBC applied to $A_\mu$, as in (\ref{eq_vac}), (\ref{eqZ}).

Eq.(\ref{5sc}) is the basis of our approach. In what follows we
shall calculate however not $F_1$, which contains all
tower of excited states over the  ground state of heavy quarks
$Q\bar Q$, but rather the static potential $V_1(r,T)$,
 corresponding to this ground state, for more details see \cite{27sc}.

 Separating, as in BPTh \cite{23sc} the field $A_\mu$ into NP
 background $B_\mu$ and valence gluon field $a_\mu$,
\be A_\mu=B_\mu +a_\mu
\label{16sc}
\ee
 one can assign gauge transformations as follows
 \be B_\mu\to U^+ (B_\mu +\frac{i}{g} \partial_\mu) U, ~~ a_\mu
 \to U^+ a_\mu U.
 \label{17sc}
 \ee
As a next step one inserts (\ref{16sc}) into (\ref{5sc}) and
expands in powers of $ga_\mu$, which gives
\be
\lan W(r, \beta) \ran =
\lan W^{(0)} (r, \beta) \ran _B + \lan W^{(2)} (r, \beta) \ran_{B,a} +\ldots,
\label{18sc}
\ee
 where according to \cite{23sc} one can   write  $\lan \Gamma
\ran_A=\lan\lan \Gamma\ran_a\ran_B$, and $\lan W^{(2)}\ran$
can be written as
\be
\lan W^{(2)}\ran_{B,a} = \frac{(ig)^2}{N_c} \int \lan \tr
P \Phi (\prod_{xy}) \lan a_\mu (x) a_\nu(y)\ran_a \Phi
(\coprod^{xy}) \ran_B dx_\mu dy_\nu.
\label{19sc}
\ee

\begin{figure}[h]
\centerline{\includegraphics[height=50mm,keepaspectratio=true]{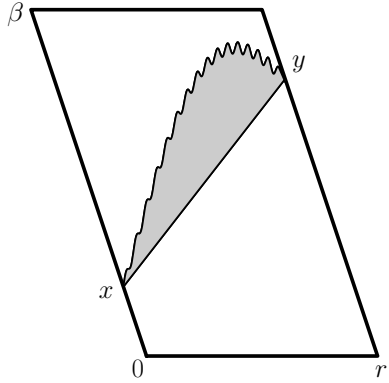}}
\caption{The gluon trajectory (wavy line) and the adjoint surface $S_{gl}^H$ (dark region)
attached to the thermal Wilson loop.}
\label{fig1}
\end{figure}

Here $\Phi (\prod)$ and $\Phi(\coprod)$  are parallel transporters
along the pieces of the original Wilson loop $W(r, \beta)$, which
result from the dissection of the Wilson loop at points $x$ and
$y$, see Fig.\ref{fig1}. Thus the Wilson loop $W^{(2)} (r, \beta)$ is the standard loop
$W^{(0)} (r, \beta)$   augmented by the adjoint line connecting points
$x$ and $y$. It is easy to see using (\ref{17sc}), that this
construction is gauge invariant.

For OGE propagator one can write the path integral Fock-Feynman-Schwinger (FFS)
representation  for nonzero $T$ as in \cite{23sc}

$$G_{\mu \nu} (x, y)= \lan a_\mu (x) a_\nu (y)\ran_a=$$
 \be
=\int^\infty_0 ds \int (D^4z)^w_{xy} \exp (-K) \Phi^{adj} (C_{xy})
\left(P_F \exp (2ig \int^s_0 F_{\sigma \rho}(z(\tau)) d \tau)\right)_{\mu \nu},
\label{20sc}
\ee
$$\Phi^{adj} (C_{xy})= P\exp (ig \int_{C_{xy}} B_\mu dz_\mu),$$
 where the open contour
$C_{xy}$ runs along the integration path in (\ref{20sc}) from the
point $x$ to the point $y$ as shown Fig.\ref{fig1},
and $K=\frac14 \int^s_0 (\dot z_\mu)^2 d\tau$.
The path integration measure $(D^4z)^w_{xy}$ is given by
\be (D^4z)_{xy}^w =\prod^N_{k=1}
   \frac{d^4\Delta z(k)}{(4\pi\varepsilon)^2}
  \int \frac{d^4p}{(2\pi)^4} \sum^{+\infty}_{n=-\infty} \exp \left({ip_\mu  \left(\sum^N_{k=1} \Delta
   z_\mu(k)-(x-y)_\mu-n\beta\delta_{\mu 4}\right)}\right)
\label{15sc}
\ee
with $N\varepsilon =s$ and $\Delta z_\mu(k) = z_\mu(k) - z_\mu(k-1)$. Thus, $(D^4z)_{xy}^w$
is a path integration with boundary conditions
$z_\mu(\tau=0) = x_\mu$ and $z_\mu(\tau=s) =y_\mu$ (this is marked by the subscript $xy$)
and with all possible windings in the Euclidean temporal direction (this is marked by the superscript $w$).

We must now average over $B_\mu$ the geometrical construction
obtained by inserting (\ref{20sc}) into (\ref{19sc}), i.e.
\be
\lan \Phi (\prod_{xy})  \Phi^{adj} (C_{xy} ) \Phi
(\coprod^{xy})\ran_B\equiv \lan W_{xy} (r,\beta)
\ran_B.
\label{21sc}
\ee
One can apply to (\ref{19sc}) the
nonabelian Stokes theorem, and to this end one has to fix the
surface bounded by the rectangular $r,\beta $ with the adjoint line
passing  on the surface. The standard prescription of the minimal
surface valid for the fixed boundary contours, in our case when chromoelectric
confinement is missing and only spatial projections
of the surface enter, leads to the  deformation of the original plane surface due to
gluon propagation, consisting of this original surface plus  the
additional adjoint surface $S_{gl}^H$ connecting gluon trajectory
with  its projection on the plane $(r,\beta)$, see Fig.\ref{fig1}, where this projection
is simplified to be the straight line.
The nonabelian  Stokes theorem yields the area law \cite{16sc, 17sc}
for distances $r \gg \lambda_g$, $\lambda_g-$ gluon correlation length, $\lambda_g \sim 0.2$ fm
\be
\lan W_{xy}(r,\beta)\ran_B = \exp  (-\sigma^E S_{\rm plane}) \exp
(-\sigma^E_{adj} S^E_{gl} -\sigma_{adj}^H S_{gl}^H),
\label{22sc}
\ee
where $S^{E,H}_{gl}$ are projections of gluon-deformed piece of
surface $S_{gl}$ into time-like, space-like surfaces respectively.

For $T>T_c$ one has $\sigma^E\equiv 0$ and one obtains exactly the
form containing the gluelump Green's function
\be
\lan W^{(2)}\ran_{B,a} ={(ig)^2} C_2(f) \int^\beta_0 dx_4 \int^\beta_0 dy_4 G_{44}(r, t_4),
 \label{13sc}
 \ee
where  $t_4\equiv x_4-y_4$, $C_2(f)=(N_c^2-1)/2N_c$ and $G_{44} (r, t_4)$ is
\be
G_{44}(r, t_4) =\int^\infty_0 ds \int (D^4z)_{xy}^w \exp (-K) \exp(-\sigma_{adj}^H S_{gl}^H).
\label{14sc}
\ee

In (\ref{14sc}) we have  neglected  the last exponent
on the r.h.s. of (\ref{20sc}), which produces spin-dependent terms found small in \cite{31sc},
for more discussion see Appendix 4 of \cite{29sc}.

Thus the gluon Green's function in the confined phase becomes a gluelump Green's function, where the adjoint
source trajectory is the projection of the gluon trajectory on the
Wilson loop plane.

Now  in  the deconfined phase, $T\geq T_c$, where, magnetic confinement takes place in
spatial coordinates, so that one can factorize as follows $(G_{\mu \nu} (x, y)\equiv \delta_{\mu \nu}G (x,y))$

\be G (x,y) = \int^\infty_0 ds
\int (Dz_4)^w_{x_4y_4}(Dz_3)_{x_3y_3}
G^{(2)} (0,0;s)\exp \left({-\frac14 \int^s_0 (\dot z^2_3 +\dot z^2_4)d\tau}\right),
\label{16}
\ee
 where  $G^{(2)} (0,0;s)$ is the 2d Green's function with
$s$ playing the role of time and interaction given by the area law
term, $\exp (-\sigma_{adj}^H S_{gl}^H)$. Here $S_{gl}^H $ is the Nambu-Goto
expression
\be
S_{gl}^H = \int^s_0 d\tau \int^1_0 d\beta \sqrt{\dot
w^2_i w_k^{'2}-(\dot w_iw'_i)^2},
\label{17}
\ee
 and $w_i= z_i(\tau)
\beta, ~~ i=1,2, ~~ w'_i =\frac{\partial w_i}{\partial\beta},
~~\dot w_i =\frac{\partial w_i}{\partial \tau}$.

In (\ref{16}) one can specify coordinates in such a way, that
$x_4=0, y_4=t_4, x_3=0, y_3=r$ and $x_{1,2}=y_{1,2}$.

For $G^{(2)}$ one can write
\be
G^{(2)}(0,0;s)=\int
(Dz_1)_{00}(Dz_2)_{00} \exp \left({-\frac14\int^s_0 (\dot z^2_1 +\dot z^2_2)
d\tau - \sigma_{adj}^H S_{gl}^H}\right).
\label{18}
\ee
The path integral (\ref{18}) can be expressed through the
 Hamiltonian $H^{(2)}$, which is obtained from the Euclidean action
 \be
 A=\int^s_0 d \tau
\mathcal{L } (z_i, \dot  z_i) =\frac14 \int^s_0 (\dot z^2_1+\dot
z^2_2) d\tau + \sigma_{adj}^H S_{gl}^H
\label{19}
\ee
\be
G^{(2)} (x,y; s) =
\lan x |\exp({-H^{(2)} s})|y\ran
\label{20}
\ee
It is easy to derive, that $G^{(2)}(0,0; s)\equiv G^{(2)}(s)$ behaves at small
and large $s$ as
\be G^{(2)} (s\to 0) \propto \frac{1}{4\pi s}; ~~ G^{(2)} (s\to \infty) \propto M^2_0 \exp ({-M^2_0 s}),
\label{21}
\ee
where $M^2_0$ is the lowest mass eigenvalue of $H^{(2)}$.
 As an explicit example one can consider a rather realistic case when interaction
 term $\sigma_{adj}^H S_{gl}^H$ in (\ref{18}) is replaced by the oscillator term
 $$\sigma_{adj}^H S_{gl}^H\to \frac14 \int^s_0 d\tau \bar \omega^2 z^2_i (\tau)$$
 and one obtains
\be
G^{(2)}_{osc} (s) =\frac{\bar
\omega}{4\pi \sinh \bar \omega s}, ~~\bar \omega=M^2_0
\label{22}
\ee

One can see that asymptotic behaviour (\ref{21}) is satisfied provided that $\bar \omega =M_0^2$.
 On the other hand, the eigenvalues  of $H^{(2)}$
 (when the role of time is played by $s$), $M_n^2$ can be expressed through
 $\tilde M^2_n$, where $\tilde M_n$ are eigenvalues of gluelump Hamiltonian  $\tilde
 H^{(2)}$ when Eucledian time evolution is chosen along $z_3$.
 Those will  be found in the next section. Since $d\tau ={dz_3}/{2\mu},~~
 dz_3\equiv dt$ and  $\tilde M_n = 2\mu_n$, the Hamiltonian $\int
 H^{(2)}d\tau = \int ({H^{(2)}}/{2\mu}) d z_3 =\int \tilde
 H^{(2)} dz_3$ and one has the equality $M^2_n \cong \tilde M^2_n$
 with the accuracy of $\sim 5\%$ known for the einbein technic
 calculations \cite{24sc}.

 Calculating $(Dz_3)_{0r}$ one has
 \be
 \int
(Dz_3)_{0r}\exp \left( -\frac14 \int^s_0 \dot z_3^2 d\tau\right)
=\frac{1}{\sqrt{4\pi s}} \exp \left(- \frac{r^2}{4s}\right).
\label{23}
\ee
 A similar calculation with $(Dz_4)^w_{ot_4}$ yields
\be
\int(Dz_4)^w_{ot_4}
\exp\left( -\frac14 \int^s_0 \dot z_4^2 d\tau\right)
=\frac{1}{\sqrt{4\pi s}} \sum^{+\infty}_{n=-\infty} \exp
\left(-\frac{(t_4+n\beta)^2}{4s}\right),
\label{24}
\ee
and combining all terms one has
 \be
 G(r,t_4)
=\frac{C_2(f)g^2}{16\pi^2} \int^\infty_0
\frac{ds}{s^2} \tilde G^{(2)}(s) \sum^{+\infty}_{n=-\infty}
\exp \left({-\frac{[(t_4+n\beta)^2 +r^2]}{4s}}\right),
\label{25}
\ee
where we have defined $\tilde G^{(2)} (s) \equiv 4\pi s G^{(2)} (0,0;s)$
so that $\tilde G^{(2)} (s\to 0) \to 1, ~~\tilde G^{(2)}
(s\to \infty) \to \exp (-M^2_0 s)$.

We are now in the position to obtain the screened static (color Coulomb) potential. Indeed, identifying
in the lowest order in $O(g^2)$ in (\ref{5sc}), (\ref{13sc}), (\ref{14sc}), (\ref{25}), one has
\be
F_1(r,\beta) =V_1^{(1)} (r,\beta) =- C_2(f) g^2 \int^{\frac{\beta}{2}}_{-\frac{\beta}{2}} d t_4 G(r, t_4),
\label{26}
\ee
which can be rewritten using (\ref{25}) as
\be
V_1^{(1)} (r,\beta) =-\frac{C_2(f) g^2}{16\pi^2} \int^\infty_0 \frac{ds}{s^2} \exp \left({-\frac{r^2}{4s}}\right)
\chi (s,\beta) \tilde G^{(2)} (s),
\label{27}
\ee
where $\chi(s,\beta)$
\be
\chi(s,\beta) =\int^{\frac{\beta}{2}}_{-\frac{\beta}{2}}
dt_4 \sum^{+\infty}_{n=-\infty} \exp \left(-\frac{(t_4+n\beta)^2}{4s}\right).
\label{28}
\ee

Now for large $\beta$ (small $T$), $\beta \gg r , \beta M_0 \gg 1$,
 one can keep in the sum (\ref{28}) only the term $n=0$, which yields $\chi_{n=0} (s,\beta) =
\sqrt{4\pi s}$. From (\ref{27}) one  then can conclude that $s\sim r^2$,
and for $r^2M^2_0 \sim sM_0^2\gg 1$ one can replace $\tilde G^{(2)} (s)$ by the
asymptotics, $\tilde G^{(2)} (s) \cong \exp (-M^2_0 s)$, which yields
\be
V_1^{(1)} (r,T) =-\frac{C_2(f)\alpha_s}{r}
e^{-M_0 r},~~ rT\ll 1.
\label{29}
\ee
In the  opposite limit of small $\beta$ (large $T$), $\beta \ll r$,
one can use the following relation \cite{dmam} for the sum in (\ref{28})
\be
\sum^{+\infty}_{n=-\infty} \exp \left(-\frac{(t_4+n\beta)^2}{4s}\right)
= { \frac{\sqrt{4\pi s}}{\beta}} \sum^{+\infty}_{k=-\infty} \exp \left({ -{\frac{4 \pi^2 k^2} {\beta^2}} s
+ i{\frac{2\pi k} {\beta}} t_4}\right),
\label{30}
\ee
which yields for $\chi(s,\beta),$
\be
\chi(s,\beta)=\sqrt{4\pi s} \left( 1+ O(e^{-{{4 \pi^2 k^2 s}/{\beta^2}}}) \right)
\label{31}
\ee
and hence the screened color Coulomb potential $V_1^{(1)} (r,T)$ has the form (\ref{29}) also at
 $rT\gg 1$. We shall assume accordingly that (\ref{29}) holds for all temperatures
and distances $r \geq \lambda_g$ in the order $O(\alpha_s)$,
 and the next section will be devoted to the calculation of $M_0$.

\section{Nonperturbative Debye mass}

As it was argued in the previous section, the screened gluon
propagator is actually the gluelump Green's function, defined in
(\ref{14sc}). In this section we shall calculate the gluelump
spectrum and hence the set of Debye masses. This problem is
similar to the calculation of the so-called meson and glueball
screening masses, which was done analytically in \cite{25sc}, and
in our present case we must compute the gluelump screening masses.
Below we shall heavily use the glueball calculation of
\cite{25sc}, simplifying it to the case, when one of the gluon
masses is going to infinity--thus yielding a gluelump.

We note, that the role of time is played by the coordinate $z_3$,
(when the third axis passes through the  positions of $Q$ and
$\bar Q$).

So we write $z_3\equiv t_3, ~~0\leq t_3\leq r$, and define
transverse vector $\vez_\perp =(z_1,z_2)$ and $z_4(t_3)$. Introducing
the einbein variable $\mu$ \cite{32sc}, one has
 \be
\frac{dz_3}{d\tau} =2\mu,~~ 0\leq\tau\leq s;~~ K=\frac12 \int^r_0
dt_3 \mu (t_3) (1+\dot \vez^2_\perp+\dot z^2_4)
\label{24sc}
\ee
and
$G(x,y)$ acquires the form
\be
G(x,y) =\int D\mu \int (D^2 z_\perp)_{00}
(Dz_4)^w_{xy} \exp (-A),
\label{25sc}
\ee
 where the action is
 \be
A=K+ \sigma^H_{adj}S^H_{gl}
\label{26sc}
\ee

Proceeding as in \cite{25sc} one arrives to the effective
Hamiltonian representation
 \be
 G(x,y) = \lan x|\sum_n
\exp({-H_n r})|y\ran,
\label{27sc}
\ee
with the temperature-dependent Hamiltonian $(r_\perp \equiv |\vez_\perp|)$
\be H_n = \sqrt{\vep^2_\perp + (2n\pi T)^2}
+\sigma_{adj}^H r_\perp.
\label{28sc}
\ee
The spatial gluelump masses are to
be found from the eigenvalues  of the equation
\be H_n
\varphi^{(n)}_k  = M^{(n)}_K \varphi_k^{(n)}.
\label{29sc}
\ee
and for $n=0$ the Hamiltonian (\ref{28sc}) has the form
\be
H_0 =
\sqrt{\vep^2_\perp} + \sigma_{adj}^H r_\perp
\label{30sc}
\ee
or in the form
with  einbein variables  which will be useful for discussion
 \be
 H_0^{\rm einb} = \frac{\vep^2_\bot}{2\mu} +\frac{\mu}{2} +\sigma_{adj}^H r_\perp=
 \frac{\vep^2_\perp}{2\mu}
+\frac{\mu}{2} +\frac{{\sigma_{adj}^H}^2 r^2_\perp}{2\nu} +\frac{\nu}{2}.
\label{31sc}
\ee
The OGE potential, $\Delta V =-3 \alpha_s^{\rm eff}/r$, will be considered as the small correction.
Note the difference between two-dimensional distance $r_\perp$
 entering in the spatial protection of the area in the gluelump Wilson loop, $S_{gl}^H$, and
 the $3d$ distance $r$ entering in the $3d$ color Coulomb interaction in $\Delta V$.
   The eigenvalue  of
(\ref{31sc}), $H_0^{\rm einb} \varphi=\varepsilon_0 \varphi$,
with $\alpha_s^{\rm eff} =0$ is
\be
\varepsilon_0 (\mu,\nu) =
\frac{\mu+\nu}{2}
+\frac{\sigma_{adj}^H}{\sqrt{\mu\nu}}
\label{32sc}
\ee
 and the minimization in $\mu,\nu$ implied in the einbein
 formalism \cite{32sc} yields
 \be
 \varepsilon_0 (\mu_0, \nu_0) = 2 \sqrt{\sigma_{adj}^H} = 3
 \sqrt{\sigma_s},
 \label{33sc}
 \ee
 where $\sigma_s$ is the fundamental spatial string tension and
 $\sigma_{adj}^H = (9/4) \sigma_s$ for $SU(3)$.

 One can compare this value with more exact one, obtained from solution of the differential
   equation in  (\ref{31sc}) and to this end one can use the eigenvalue of
 the screening glueball mass found in \cite{25sc}, (which  is larger by a
 factor of $\sqrt{2}$ than that of our gluelump mass, cf. Eq.
 (\ref{35sc}) of \cite{25sc} and our Eq. (\ref{30sc})). In this way
 one obtains
 \be
 \varepsilon_0=2.82 \sqrt{\sigma_s}\label{34sc}\ee which
 differs from (\ref{33sc}) by 6\%.

 In the next approximation  the OGE potential
 for the gluelump comes into play. Here one should take into account that the
 gluon-gluon OGE interaction acquires a large  NLO correction, which strongly reduces the LO result  as
 it is seen in the BFKL calculation  (see discussion in \cite{33sc}), and therefore the
 effective value of $\alpha_s^{\rm eff}$ is smaller than in the  $Q\bar Q$
 interaction. Specifically, in the gluelump mass calculation at
 $T=0$ \cite{31sc} the mass of the lowest gluelump for $\alpha_s^{\rm eff} =0$ is $M=1.4$ GeV,
 and  it decreases to $M\approx 1$ GeV, when $\alpha_s^{\rm eff} =0.15$.
 This latter value of $M$ is in agreement with
 lattice correlator calculations
 \cite{21sc};  the same situation takes place  in the glueball
 mass calculation \cite{33sc}, where also $\alpha_s^{\rm eff} \cong 0.15$
 and we shall adopt it in our  Eq. (\ref{31sc}). The
 correction of  $\varepsilon_0$ due to  $\alpha_s^{\rm eff}$ in
 the lowest order is easily computed using (\ref{31sc});  as a result one has
  $\Delta\varepsilon_0=- (9/\sqrt{\pi}) \alpha_s^{\rm eff}
 \sqrt{\sigma_s}\approx -5.08 \alpha_s^{\rm eff} \sqrt{\sigma_s}$.
 As a final result we write  the Debye mass (lowest gluelump mass $ M_0\equiv m_D$) for
 $\alpha_s^{\rm eff} =0.15$
 \be
 m_D= \varepsilon_0 +\Delta\varepsilon_0 =
 (2.82 - 5.08 \alpha_s^{\rm eff}) \sqrt{\sigma_s} \cong 2.06  \sqrt{\sigma_s}
 \label{35sc}
 \ee

\section{Numerical results and discussion}

One can now compare our prediction for $m_D(T) =c_D
\sqrt{\sigma_s(T)}$ with the latest lattice data \cite{15sc}. The
spatial string tension is chosen in the form \cite{26sc,38sc}
\footnote{Physical justification for
resorting to dimensionally reduced regime at $T=\sqrt{\sigma_s(T)}$
was given by \cite{22sc} (see also \cite{Agasian:1997wv})}

\be
\sqrt{\sigma_s(T)}= c_\sigma g^2 (T) T,
\label{sigmas}
\ee
with the two-loop expression for $g^2(T)$
\be
\label{coupl}
g^{-2}(t) =2 b_0 \ln \frac{t}{L_\sigma}+
\frac{b_1}{b_0}\ln \left(2\ln \frac{t}{L_\sigma}\right),~~~t\equiv \frac{T}{T_c}
\ee
where
$$
b_0=\left(\frac{11}{3}N_c-\frac{2}{3}N_f \right)\frac{1}{16\pi^2},~~~
b_1=\left(\frac{34}{3}N_c^2-(\frac{13}{3}N_c-\frac{1}{N_c})N_f\right)\frac{1}{(16\pi^2)^2}.
$$

\begin{figure}[!ht]
\begin{picture}(440,185)
\put(20,35){\includegraphics[width=210pt,height=140pt]{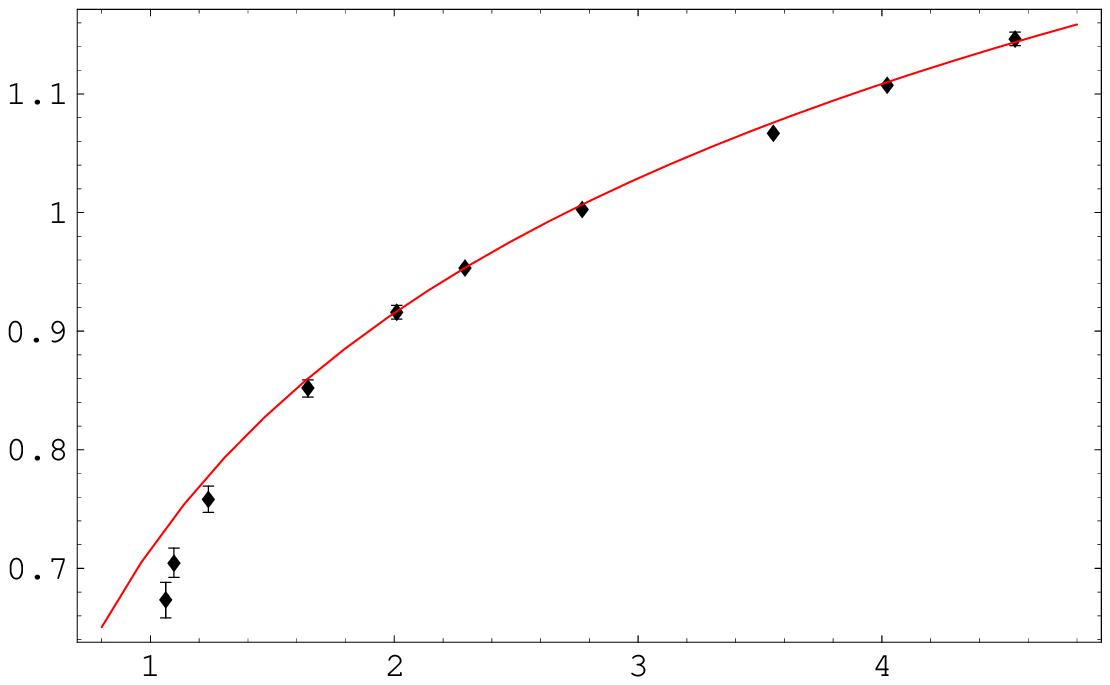}}
\put(0,90){\rotatebox{90}{$T/\sqrt{\sigma_s(T)}$}}
\put(240,35){\includegraphics[width=210pt,height=140pt]{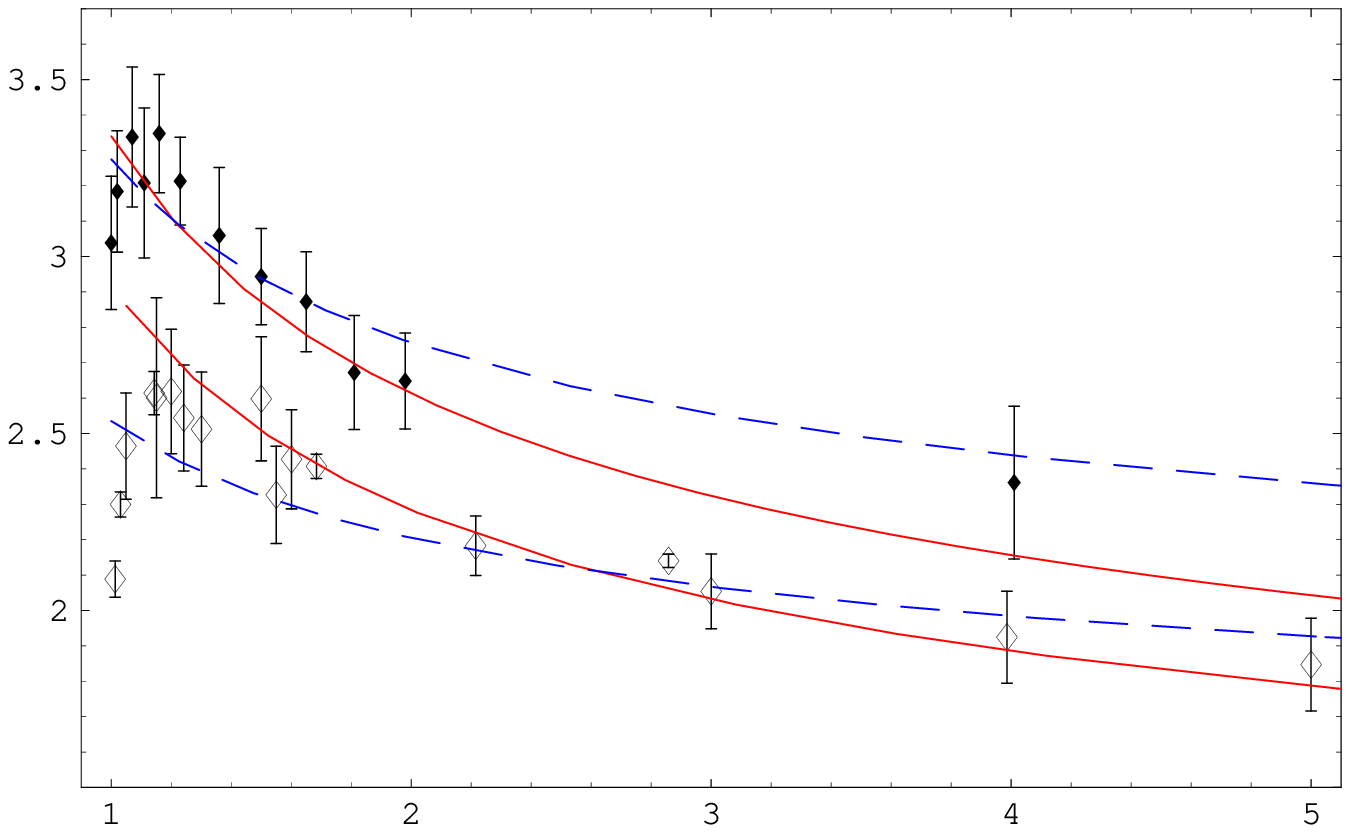}}
\put(455,90){\rotatebox{90}{$m_D(T)/T$}} \put(200,20){$T/T_c$}
\put(430,20){$T/T_c$}
\end{picture}
\caption{Left panel: The temperature over the square root of the spatial string tension
versus $T/T_c$ for pure glue  QCD. Solid line corresponds to Eq.(\ref{sigmas}).
The lattice data are from \cite{26sc}.
Right panel: Chromoelectric Debye mass $m_D/T$ for 2-flavor QCD (upper lines)
and quenched ($N_f=0$) QCD (lower lines)
versus $T/T_c$. Solid lines are calculated using
$m_D(T) =2.09 \sqrt{\sigma_s(T)}$, where $\sqrt{\sigma_s(T)}$ corresponds
to Eq.(\ref{sigmas}) with $N_f=2$ for upper solid line and $N_f=0$ for lower solid line.
Dashed lines are calculated using Eq.(\ref{37sc}), $N_f=2$--upper line, $N_f=0$--lower line.
The lattice data are from \cite{13sc}.}
\label{fig_mdeb}
\end{figure}

The measured in \cite{26sc} spatial string tension in pure glue QCD
corresponds to the values of
$c_\sigma=0.566\pm 0.013$ and  $L_\sigma\equiv \Lambda_\sigma/T_c=0.104\pm 0.009.$
On the left panel of Fig.~\ref{fig_mdeb} are shown lattice data \cite{26sc}
and the theoretical curve (solid line) for $T/\sqrt{\sigma_s(T)}$
calculated according to (\ref{sigmas}) with
$c_\sigma=0.564, L_\sigma=0.104$ and $N_f=0$.

On the right panel of Fig.~\ref{fig_mdeb} are shown lattice data \cite{13sc} and theoretical curves
for the Debye mass in quenched ($N_f=0$) and 2-flavor QCD. Solid lines correspond to our
theoretical prediction, $m_D(T) =c_D \sqrt{\sigma_s(T)}$, with $c_D=2.09$
and for $\sqrt{\sigma_s(T)}$ we exploit the same parameter
($c_\sigma=0.564, L_\sigma=0.104$) as in the left panel. The upper solid line is for
the Debye mass in 2-flavor QCD, and the lower -- for quenched QCD. We note that in computing
$m_D(T)$ using (\ref{sigmas}), (\ref{coupl})
all dependence on $N_f$ enters only through the Gell-Mann--Low coefficients $b_0$ and $b_1$.
For comparison we display in the right panel of Fig.~\ref{fig_mdeb} dashed lines for  $m_D(T)/T$,
calculated with a perturbative inspired ansatz \cite{15sc}

\be
m_D^{Latt}(T)= A_{N_f} \sqrt{1+\frac{N_f}{6}} g(T) T.
\label{37sc}
\ee
Quenched QCD corresponds to $A_{N_f=0} = 1.51$ and $L_{\sigma}^{N_f=0}=1/(1.14\cdot2\pi)$ \cite{15sc},
and for the 2-flavor QCD  $A_{N_f=2} = 1.42$ and $L_{\sigma}^{N_f=2}=1/(0.77\cdot2\pi)$ \cite{15sc}.

Let us now consider higher orders of BPTh for $m_D$. From the
gauge-invariant expansion (\ref{18sc}) one obtains the next term
$\lan W^{(4)}(r,\beta)\ran$, which contains the double gluon
propagator $\lan a_{\mu_1} (x_1) a_{\mu_2} (x_2) a_{\nu_1} (y_1)
a_{\nu_2}(y_2)\ran_a$ in the background field of the Wilson loop,
which is proportional to $g^4(T)$. One can show that the background averaging of this
propagator attached to the Wilson loop  yields the diagram of the
exchange of a double gluon gluelump between $Q$ and $\bar Q$, and
therefore the NLO BPTh Debye mass will coincide with the double
gluon gluelump mass, computed for $T=0$ in \cite{31sc}
analytically and in \cite{39sc} on the lattice. As a result the
lightest $2g$ gluelump mass appeared to be 1.75 times heavier than
the lightest $1g$ gluelump mass. We expect therefore that also at
$T>T_c$ the  same ratio of masses takes place, so that the
asymptotics of gluon (gauge-invariant and background-averaged)
gluon exchanges in BPTh has the form \cite{23sc, 40sc}
 \be
 V^{GE} (r,T) =- \frac43 \frac{\alpha_s^{(0)}}{r}
 e^{-m_D^{(1gl)}(T)r} -\frac{c_2(r) (\alpha_s^{(0)})^2}{r}
e^{-m_D^{(2gl)}(T) r}+\ldots,
\label{40sc}
\ee
where $c_2(r)$ contains the asymptotic freedom logarithm.

\begin{figure}[h]
\centerline{\includegraphics[height=50mm,keepaspectratio=true]{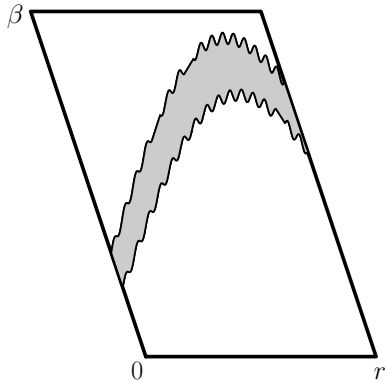}}
\caption{NLO BPTh contribution to the Debye screening potential from the $2g$ glueball exchange}
\label{fig2m}
\end{figure}

One can see in (\ref{40sc}) that the second term on the r.h.s. is
subleading and small as  compared to the first one, both due  to
$(\alpha_s^{(0)})^2$ and due to higher mass of $m_D^{(2gl)}$.
 At this point it is essential to note that this second   term
 should enter as a sum over all possible $2g$ gluelumps. In one
 particular case, when two gluons form a color singlet, they
 decouple from the plane surface of the Wilson loop and create a
 $2g$ glueball, coupled by the spatial string (see Fig.\ref{fig2m}). The corresponding
 glueball mass is computed in \cite{25sc} and is 1.7 times larger
 than the LO BPTh mass (\ref{35sc}), which is denoted as
 $m_D^{(1gl)}(T)$ in  (\ref{40sc}). It is interesting that the aforementioned
 glueball mass  corresponds to the gauge-invariant
 Debye mass suggested in \cite{34sc}, and as  seen in
 (\ref{40sc}) it appears in the NLO BPTh,  giving a small
 correction to the LO Debye screening potential. Therefore one can identify
 the Debye mass $m_D \equiv m_D^{(1gl)}$ with the accuracy
 $O(\alpha_s^{(0)} e^{-\Delta m_D r})$ where $\Delta m_D= m_D^{(2gl)}-m_D^{(1gl)} \geq 0.6 GeV$.

 The consideration above was done for the chromoelectric Debye
 mass, which appears in the screening Coulomb potential in the
 temporal plane $(4i)$ appearing due to the $G_{44} (x,y)$
 gluelump Green's function.  One can similarly consider exchange of "magnetic  gluon" by insertion
  of the magnetic field vertex $F_{ik} \sim D_i a_k -D_k a_i$ into the Wilson or Polyakov loops.
  This vertex automatically appears in the Green's function from the term   $\propto \exp (g\sigma_{\mu\nu}
  \int^s_0 F_{\mu\nu} d\tau)$ creating spin-dependent interaction. The
  same procedure as above leads to the "magnetic gluelump"
  Green's function, which differs from the "electric gluelump" case by the
  nonzero gluelump momentum $L=1$. The  corresponding mass is easily obtained as in (\ref{34sc}), giving
  $$\varepsilon_L (\mu_0, \nu_0) =2 \sqrt{(1+L+2n_r) \sigma_{adj}},
  ~~ \varepsilon_1 =\sqrt{2} \varepsilon_0 \approx 4\sqrt{\sigma_s}.$$
  Thus nonperturbative magnetic Debye mass is $\sqrt{2}$ times heavier than the electric one.

\section{Conclusions}

We have studied Debye screening in the hot nonabelian theory. For that purpose
the gauge-invariant definition of the free energy of the static
$Q\bar Q$-pair in the singlet color state was given in terms of the thermal Wilson loop.
Due to the chromomagnetic confinement persisting at all temperatures $T$, the hot QCD is
essentially nonperturbative. To account for this fact in a gauge-invariant way the BPTh was
developed for the thermal Wilson loop using path-integral FFS formalism. As a result one obtains
from the thermal Wilson loop the screened Coulomb potential with the screening mass corresponding
to the lowest gluelamp mass. Applying the Hamiltonian formalism to the BPTh Green's functions
with the einbein technic the gluelump mass spectrum was obtained. As a result, we have
derived the leading term of the BPTh for the Debye mass which is the purely nonperturbative,
$m_D(T)=c_D\sqrt{\sigma_s(T)}$ with $c_D \approx 2.06$.

Comparison of our theoretical prediction (solid lines on the right panel Fig.~\ref{fig_mdeb})
with the perturbative-like ansatz (\ref{37sc}) (dashed lines) shows that both agree
reasonably with lattice data in the temperature interval $T_c < T \leq 5 T_c$;
the agreement is slightly better for our results.
At the same time, in (\ref{37sc}) a fitting constant is used $A_{N_f}\sim 1.5$, which
is necessary even at $T/T_c\sim 5$. At this point one can discuss the accuracy
and approximations of our approach. As it was checked in numerous applications
to hadron masses and wave function (see review \cite{17sc})
the accuracy of the Hamiltonian technic is around $(5 \div7)$\%, while the area law is as
accurate for loop sizes beyond $\lambda_g \sim 0.2$ fm. At smaller distances the area law
in (\ref{14sc}), (\ref{18}) is replaced by the "area squared" expression \cite{16sc}
which yields effectively much smaller $m_D(T)$. Therefore we expect that the Debye regime
(\ref{1sc}) with the $m_D$ as in (\ref{35sc}) starts at $r \geq \lambda_g \approx 0.2$ fm.
As a whole we expect the accuracy of the first approximation of our
approach, Eq.~(\ref{35sc}) to be better than 10\%, taking also into account the bias in the definition
of $\alpha_s^{\rm eff}$ for the gluelump. The temperature region near $T_c$ needs
additional care because i) the behaviour (\ref{sigmas}) deviates from the data
(see left panel of Fig.~\ref{fig_mdeb}) and ii) contribution of chromoelectric fields above
$T_c$ (correlator $D_1^E$, see \cite{27sc}) which was neglected above.
Both points can be cured and will be given elsewhere.

The authors are grateful to Frithjof Karsch
for supplying us with numerical data, useful remarks and  fruitful discussion
with one of the authors (Yu.S).
Our gratitude is also to M.A.Trusov for help and useful advices.

The work  is supported
by the Federal Program of the Russian Ministry of Industry, Science, and Technology
No.40.052.1.1.1112, and by the grant  of RFBR No. 06-02-17012,
and by the grant for scientific schools NS-843.2006.2.

\end{document}